\documentclass[runningheads]{llncs}
\usepackage[T1]{fontenc}

\usepackage{graphicx}
\usepackage[table]{xcolor}
\usepackage{amsmath}
\usepackage{amsfonts}

\usepackage{tabularx}
\newcolumntype{Y}{>{\centering\arraybackslash}X}

\usepackage{diagbox}

\begin{document}
\title{BraSyn 2023 challenge: Missing MRI synthesis and the effect of different learning objectives}
\titlerunning{MRI sequence synthesis}
\author{Ivo M. Baltruschat \and
Parvaneh Janbakhshi \and
Matthias Lenga}
\authorrunning{I. M. Baltruschat et al.}
\institute{Bayer AG, Müllerstr. 178, 13353 Berlin, Germany\\
\email{\{ivo.baltruschat,parvaneh.janbakhshi,matthias.lenga\}@bayer.com}}
\maketitle
\begin{abstract}
This work addresses the Brain Magnetic Resonance Image Synthesis for Tumor Segmentation (BraSyn) challenge, which was hosted as part of the Brain Tumor Segmentation (BraTS) challenge in 2023. In this challenge, researchers are invited to synthesize a missing magnetic resonance image sequence, given other available sequences, to facilitate tumor segmentation pipelines trained on complete sets of image sequences. This problem can be tackled using deep learning within the framework of paired image-to-image translation. In this study, we propose investigating the effectiveness of a commonly used deep learning framework, such as Pix2Pix, trained under the supervision of different image-quality loss functions. Our results indicate that the use of different loss functions significantly affects the synthesis quality. We systematically study the impact of various loss functions in the multi-sequence MR image synthesis setting of the BraSyn challenge. Furthermore, we demonstrate how image synthesis performance can be optimized by combining different learning objectives beneficially.
\end{abstract}

\section{Introduction}

Automatic localization and segmentation of brain tumors in magnetic resonance imaging (MRI) has been an emerging research area that aims to provide clinicians with efficient and objective aid in diagnosing and monitoring patients.
Tumor biological properties can be captured differently depending on the MRI sequence. Therefore, many of the recent deep learning-based segmentation algorithms require multiple input MRI sequences during the inference stage, e.g., typically T1-weighted images with and without contrast enhancement (T1-N and T1-C, respectively), T2-Weighted (T2-W) images, and T2-FLAIR (T2-F) images. However, the challenge in such multi-modal approaches arises when an MR sequence is missing due to time constraints and/or motion artifacts.

The \emph{Brain MR Image Synthesis for Tumor Segmentation} (BraSyn) challenge as part of the Brain Tumor Segmentation (BraTS) challenge 2023 provides an opportunity for researchers to address the problem of missing MRI sequence by synthesizing it given multiple available MRI sequences~\cite{Bran2023}. Therefore, researchers are invited to work on MRI synthesis or MRI image-to-image translation algorithms, the solution of which can later facilitate automatic brain tumor segmentation pipelines where the missing sequence can be substituted by its synthesized counterpart~\cite{Bran2023}.
This work focuses on the BraSyn challenge, where the specific goal is to synthesize one missing MRI sequence given the other three available sequences. The synthesized MRI scans should be perceptually as similar as possible to the missing sequences and provide the necessary information (as in the real missing scans) for a downstream task such as tumor segmentation.

The problem lies in the scope of paired image-to-image translation for either converting MRI sequences to one another or producing a Gadolinium-based contrast agent-enhanced post-contrast T1 image from pre-contrast MRI scans. Most recent image synthesis approaches exploit Pix2Pix~\cite{Isola2017}, i.e., U-Net style encoder-decoder architecture trained using different penalties, i.e., loss functions, to encourage appropriate similarities between real and synthesized images.
In~\cite{Yu2019}, a Pix2Pix model is used to synthesize T2-F and T2-W images from T1-w images by incorporating edge information into the typical loss functions used in Pix2Pix (i.e., adversarial and L1 norm) to enforce the synthesized image to have a similar edge map as the real image.
In~\cite{Liu2023}, the missing data imputation is formulated as a sequence-to-sequence prediction problem, where a sequence of (available) input sequences of variable length can be converted to the sequence of (missing) output sequences using transformer models trained based on a combination of L1 norm and adversarial loss functions.

In the literature, image-to-image translation approaches based on the Pix2Pix model, despite their simplicity, have shown promising performance in many applications. However, depending on the synthesis task, the networks have been trained under the supervision of a wide range of image-quality loss functions, which makes it difficult to interpret the contributions of each proposed architecture variant decoupled from the effects of the different training objectives.

Addressing the BraSyn challenge, in this work, we proposed to investigate the effectiveness of a commonly used deep learning image-to-image translation approach, such as Pix2Pix trained under the supervision of different image-quality loss functions. The right choice of loss function dictates the quality of synthetic images and can be crucial for the convergence of the network, in particular in scenarios with limited availability of training data. We aim to establish a baseline framework for the task at hand while providing a comprehensive comparison and benchmarking of different training procedures for brain MRI sequence synthesis (which is lacking in the literature) and validating the effectiveness of each against the evaluation scenarios considered by the challenge organizers.

We aim to investigate the contributions of various loss functions to the synthesis of realistic brain MRI images. Specifically, we consider loss functions based on pixel-to-pixel similarities (e.g., $\mathcal{L}_1$ norm and its variants), adversarial training, Structural Similarity Index (SSIM), frequency domain consistency, and latent feature (VGG-based perceptual) consistency. Experimentally, we will verify which loss function guides the network to produce more realistic images. Furthermore, we aim to investigate whether there are any differences among the synthesized sequences with respect to the loss functions used. By combining different loss functions, we demonstrate the possibility of achieving more optimal image synthesis performance.

\section{Method}
\begin{figure}
\includegraphics[width=\textwidth]{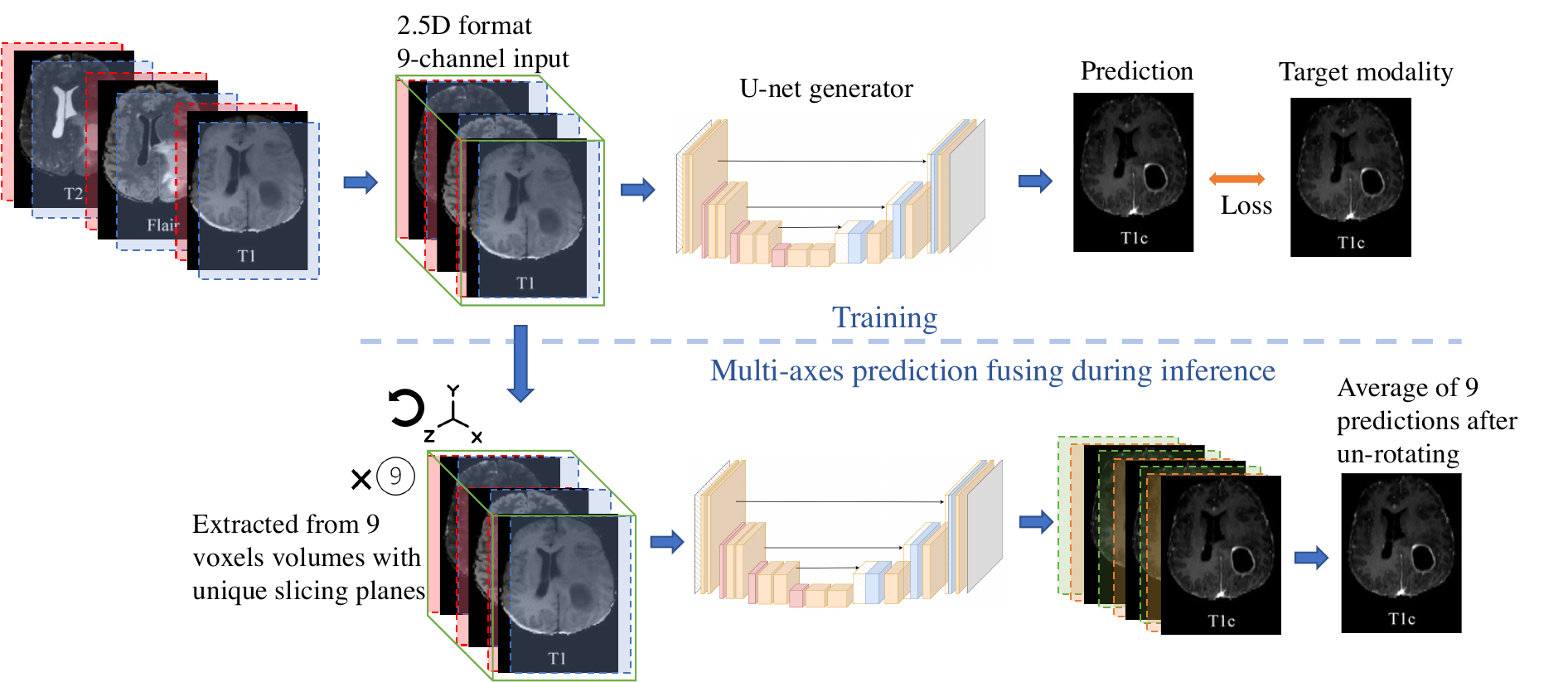}
\caption{The used synthesis framework for predicting an exemplary target sequence, e.g., T1-C from input sequences, i.e., T1-N, T2-W, and T2-F.} \label{fig1}
\end{figure}

This section describes the dataset, the synthesis framework, i.e., the network architecture, its training procedure based on different loss functions, and the inference procedure. Our general synthesis framework, i.e., training and inference stages, are depicted in Fig.~\ref{fig1}.

\subsection{Dataset}\label{data}
The BraSyn-2023 dataset is based on the RSNA-ASNR-MICCAI BraTS 2021 dataset~\cite{Bran2023}, which includes collections of brain tumor MRI scans from various institutions and annotated tumor subregions.
For training data, $1251$ scans with four complete image sequences along with their respective segmentation labels are available, i.e., i) pre-contrast T1-weighted (T1-N), ii) post-contrast T1-weighted (T1-C), iii) T2-Weighted (T2-W), and iv) T2 fluid-attenuated inversion recovery (T2-F). In the validation and test sets (including $219$ and $570$ scans, respectively), a single sequence out of the four sequences will be randomly omitted with the objective of being synthesized. All BraSyn scans have undergone standardized pre-processing according to~\cite{Bran2023}.

\subsection{Networks}
\subsubsection{Synthesis (generator) network}
Due to promising results of U-net style networks in literature~\cite{Baltruschat2023}, we adapted an architecture from~\cite{ronneberger2015unet} where we used $8$ layers of $3\times 3$ convolutions with stride $2$, each followed by Mish activation function and finally a CeLU activation function for the output layer.

Assuming all MRI sequences are co-registered and after applying appropriate normalization (as detailed in Section~\ref{training}), we adapted a multi-slice training procedure in which the input images are arranged in the format of a stack of $2.5$D slices, along either the axial, sagittal, or coronal planes (cf. Section~\ref{training}). $2.5$D input slices, unlike processing $2$D slices, provide some additional spatial information and context from the $3$D brain. To create the $2.5$D input images for each "slice of interest" in all three available MRI sequences, three consecutive slices (providing context information from two neighboring slices of the "slice of interest") are considered. After channel-wise stacking, this forms a 9-channel input image. We added zero slices for regions outside of the original volume for edge slices. The $2.5$D stack of the three available MRI sequences is fed into the generator to synthesize the corresponding central slice, i.e., the "slice of interest," from the missing sequence (i.e., target). Therefore, we trained four separate networks to synthesize each missing sequence (T1-N, T1-C, T2-W, or T2-F scans).

Models trained using $2.5$D image stacks do not fully account for the continuity in a 3D MRI volume. This can lead to specific synthesizing artifacts, which can be spotted by local structure inconsistency or global brightness changes. To alleviate the discontinuity problem in the output of our models, during inference, we incorporated the multi-axes prediction fusing approach mentioned in~\ref{MPF}. Also, unlike models operating on full $3$D volumes, they need less GPU memory and have more training data available.

We denote the mapping from the input domain $X$ to the target domain $Y$ using the synthesis or generator model by $G:X\rightarrow Y$, where the synthesized output $\hat{y}=G(x)$, $x\in R^{K_1\times K_2 \times 9}$ and $\hat{y}, y\in R^{K_1\times K_2}$, and $K_1$ and $K_2$ are dimensions of image slices.

\subsubsection{Discriminator}
In the case of using adversarial training, we also adopted a patch-wise discriminator from PatchGAN in~\cite{Isola2017} where we used a $5$ layer discriminator with additional series of spectral normalization~\cite{Miyato2018} to stabilize the training further. The discriminator is trained with the mean square loss function. As the discriminator is meant to distinguish between the synthesized and real images of the target, it takes two input slices of the target and synthesized images. We denote the discriminator function as $D$.

\subsection{Training}\label{training}
Being provided by a training set for the challenge (cf. Section~\ref{data}), we further split the training set into {\it train} subset of $1125$ scan series for model training and {\it development} subset of $126$ scan series for monitoring the training.
We applied a histogram standardization~\cite{Nyul2000} on all scans. For each sequence, the training set is used to find the intensity landmarks. Furthermore, on each set of input voxel series (i.e., $3$ voxels) and the target (to be predicted) voxel, MinMax intensity scaling has also been applied to re-scale the range of the input voxel sets and target voxels to intensity range of $[0,1]$ separately. During the inference, the normalization scale applied to the input series is used to re-scale the normalized predicted scans to the original intensity range, i.e., $[0, \inf]$.

To augment the training data, we sliced the scans in axial, sagittal, and coronal planes, which, after removing all-zero slices, resulted in $502,971$ {\it training} and $56,270$ {\it development} slices. Furthermore, we also applied online random cropping to $256\times256$ (after initially zero-padding all slices to $288\times288$), random horizontal flipping (with a probability of $p=0.5$), and random rotations in the range of $[-15, 15]$ degrees (with $p=0.5$).

Our models are trained using the ADAM optimizer with $beta_1=0.5$, $beta_2=0.99$, an initial learning rate of $0.0001$, and a batch size of $64$. We reduced the learning rate by a factor of two every $10$ epochs and trained the networks for a total of $100$ epochs, where $400,000$ training images are used in each epoch.

In this work, we investigated the impact of different synthesis loss functions to train networks to predict each missing sequence, where, in the following, the considered loss functions are briefly introduced.

\subsubsection{L1 loss}
L1 norm is the commonly used synthesis loss function in which the mean absolute error between the synthesized and target images is computed with the assumption that the images are well (pixel-to-pixel) aligned.
\begin{equation}
\mathcal{L}_1 = \mathbb{E}_{x,y}|y-G(x)| \label{l1}
\end{equation}

\subsubsection{Masked L1 loss}
$\mathcal{L}_1$ is being computed globally on the whole image (cf. eq~\ref{l1}), while one of the problems in brain tumor segmentation is that lesions affect a small portion of the brain. Therefore, for a more precise synthesis of tumor regions, the $\mathcal{L}_1$ can be modified to penalize errors in tumor regions and healthy regions of the brain separately, since the segmentation mask for each input series is also available during training. Let $M_x^t$ denote a binary mask around the tumor for each input $x$. After multiplying $G(x)$ and $y$ by $M_x^t$ and ${M_x^h}=1-M_x^t$, respectively, the corresponding masked tumor and healthy regions are denoted by $G(x)^t$ and $y^t$ and $G(x)^h$ and $y^h$ giving the following loss terms for each region:
\begin{align}
\mathcal{L}_1^{t} &= \mathbb{E}_{x,y}|y^t-G(x)^t|/|M_x^t|,\\
\mathcal{L}_1^{h} &= \mathbb{E}_{x,y}|y^h-G(x)^h|/|{M_x^h}|,\\
\mathcal{L}_1^{M}  &= w \mathcal{L}_1^{t} + (1-w) \mathcal{L}_1^{h},
\end{align}
where $w$ is the weighting factor to control the contributions of loss terms corresponding to tumor and healthy regions. Here, we used $w=0.5$.

\subsubsection{Adversarial loss}
Adversarial loss has been widely used to improve the overall perceptual quality of synthesized images by overcoming the blurriness produced by L1 norm and capturing high-frequency structures~\cite{Isola2017}.
Adversarial training is achieved through the min-max optimization objective as in~\cite{Mao2017}, where, in practice, the optimal parameters of G and D are approximated using an alternating training procedure. Motivated by the improvement of training stability, in our experiments, we used adversarial training in the least squares GAN (LSGAN)~\cite{Mao2017} framework resulting in the loss functions:
\begin{align}
\mathcal{L}_{adv}^{D} &= \mathbb{E}_{y}{(D(y)-1)}^2 + \mathbb{E}_{x}{(D(G(x))}^2,\\
\mathcal{L}_{adv}^{G} &= \mathbb{E}_{x}{(D(G(x)-1)}^2
\end{align}

\subsubsection{SSIM loss}
Motivated by considering image structures in the loss function (unlike pixel-wise criterion such as L1 loss) and improving the perceptual quality of synthesized images, SSIM loss has also been introduced and exploited for similar tasks in~\cite{Zhao2017}:
\begin{align}
\mathcal{L}_{SSIM} = \mathbb{E}_{x,y}|1-SSIM(y,G(x))|
\end{align}
Therefore, in this work, the effectiveness of SSIM loss has also been investigated. For our experiments, we set the kernel size to $11$ required for SSIM computation~\cite{Zhao2017}.

\subsubsection{Perceptual VGG loss}
  We also investigated a new variant of the perceptual loss~\cite{Simonyan2014}, the VGG conv-based perceptual loss~\cite{Zhou2022}, which has been reported to be effective in enhancing image synthesis perceptual quality. Here, we used the pre-trained VGG-19 version, and the perceptual loss is defined as
\begin{align*}
  \mathcal{L}_{VGG} = \sum _{l \in \{2, 7, 14, 21, 28\}} | \lambda _{l} \left [{ \phi _{l}^{\mathrm {conv}}\left (G(x)\right) - \phi _{l}^{\mathrm {conv}}\left (y\right) }\right] |^{2}, \tag{3}
\end{align*}
  where $G(x)$ and $y$ are the synthesized and target image, respectively. $\phi _{l}^{\mathrm {conv}}$ denotes the feature maps of the $l$-th convolutional layer of the pre-trained VGG-19. $\lambda_l$ are set to be 0.0002, 0.0001, 0.0001, 0.0002, and 0.0005.

\subsubsection{Frequency loss}
Motivated by~\cite{Baltruschat2023}, we also investigated the frequency-based training objective to enhance the image translation process by directly regulating the consistency of information in low and high-frequency domains:
\begin{align}
\mathcal{L}_{Freq} =& \mathbb{E}_{x,y}||\mathcal{F}(y)|\odot M_r-|\mathcal{F}(G(x))|\odot M_r| \\\notag &+\mathbb{E}_{x,y}||\mathcal{F}(y)|\odot \overline{M_r}-|\mathcal{F}(G(x))|\odot \overline{M_r}|,
\end{align}
where $|\mathcal{F}(.)|$ denotes the magnitude of discrete Fourier transform. $M_r$ denotes a binary mask such that the circular region around the origin with radius $r$ is set to $1$ (capturing low-frequency information) while its inverse (capturing high-frequency information) is denoted as $\overline{M_r} := 1 -  M_r$ with $\odot$ denoting the Hadamard product.
Intuitively, the low-frequency components of the loss function are meant to dictate the consistency of information, such as image brightness. On the contrary, the high-frequency components correspond to consistency in sharp edges and more fine-grained details of images~\cite{Baltruschat2023}. Similarly to~\cite{Baltruschat2023}, we set $r=21$ in our experiments.

\subsection{Inference} \label{MPF}
Our models synthesize slices of MRI images independently of each other. Therefore, after concatenating the slices to form the 3D volume, strong discontinuities may be observed along different axes, which impairs the entire image synthesis quality.
Therefore, we adapt the multi-axes prediction fusing from~\cite{baltruschat2023uncertainty,Baltruschat2021} during inference. In multi-axes prediction fusing, the input volume is reformatted into three principal axes (having three different slicing planes) where each is rotated (using interpolation) along two of the principal axes (by $45$ degree) to produce overall $9$ volumes with unique slicing planes. The final synthesized image is computed by taking the mean of the $9$ reformatted volumes after rotating back to the original acquisition plane~\cite{Baltruschat2021}.

\section{Results and Discussion}
This section describes the preliminary results obtained using synthesis networks trained under the supervision of multiple losses for different sequence synthesis tasks. Our experiments are structured as follows. First, we investigated the effect of $\mathcal{L}_1$ and $\mathcal{L}_1^{M}$ on the synthesizing of each sequence. Secondly, we combined $\mathcal{L}_1^{M}$ with each of the more advanced loss functions such as $\mathcal{L}_{adv}$, $\mathcal{L}_{SSIM}$, $\mathcal{L}_{VGG}$, and $\mathcal{L}_{Freq}$. Finally, we trained on a tuned combined loss $\mathcal{L}_{\mathrm{combined}} = 5\mathcal{L}_1^{M} + \mathcal{L}_{adv} + \mathcal{L}_{SSIM} + \mathcal{L}_{VGG} + \mathcal{L}_{Freq}$

For the evaluation of the quality of the synthesized images for each of the four tasks, similar to the evaluation metric used by the challenge organizers, we used the calculated SSIM and PSNR scores in the tumor (denoted as SSIM$^t$ and PSNR$^t$, respectively) and healthy areas of the brain (denoted as SSIM$^h$ and PSNR$^h$, respectively). All models are evaluated on our {\it development} set with 126 scans (which were not used in training).
Since the validation phase of the challenge was closed during the submission of the article (because of multiple timeline changes), we cannot report segmentation dice scores or SSIM scores on the validation set. In Section~\ref{sec:test_results}, we report the final test results of the challenge organizers. Due to time constraints, our final submission to the challenge included four models (i.e., one model for each type of sequence) that were trained with $\mathcal{L}_{\text{combined}}$.

\subsection{Training with different loss functions}
In Table~\ref{tab1} the results of training with $\mathcal{L}_1$ or $\mathcal{L}_1^{M}$ for each sequence are shown. For T1-synN and T2-synW, $\mathcal{L}_1$ performed constantly better than $\mathcal{L}_1^{M}$ in all four measurements. For T2-synF, $\mathcal{L}_1$ slightly exceeds $\mathcal{L}_1^{M}$ based on SSIM, but for PSNR it is the other way around. Finally, T1-synC performed better on three out of four scores when trained with $\mathcal{L}_1^{M}$. We chose for our second evaluation to combine $\mathcal{L}_1^{M}$ with the more advanced loss function because it helped to improve the results for the tumor regions. The results in Table~\ref{tab2} show that training a model with $\mathcal{L}_1^{M} + \mathcal{L}_{Freq}$ helped the most to improve our results for T1-synN, T1-synC, and T1-synW. Only for T2-synF, we can see that the combination of $\mathcal{L}_1^{M} + \mathcal{L}_{adv}$ has the highest overall score.

\begin{table}[th!]
\centering
\caption{Image quality results for training models (i.e., synthesizing T1-synN, T1-synC, T2-synF, or T2-synW) with $\mathcal{L}_1$ or $\mathcal{L}_1^{M}$, The input to each model are always the other three sequences, e.g., for T1-synN, the input contains information from T1-C, T2-W, and T2-F. Bold text highlights the best result for each sequence.}\label{tab1}
\begin{tabularx}{0.9\linewidth}{Y rr|rr|rr|rr}
\hline
 \multicolumn{1}{l}{}                & \multicolumn{2}{c}{T1-synN} & \multicolumn{2}{c}{T1-synC} & \multicolumn{2}{c}{T2-synW} & \multicolumn{2}{c}{T2-synF} \\ \hline
\diagbox{Metric}{Loss func.} & $\mathcal{L}_1$ & $\mathcal{L}_1^{M}$ & $\mathcal{L}_1$ & $\mathcal{L}_1^{M}$ & $\mathcal{L}_1$ & $\mathcal{L}_1^{M}$ & $\mathcal{L}_1$ & $\mathcal{L}_1^{M}$ \\
\hline\hline
 SSIM$^h$ & \textbf{0.740} & 0.723 & \textbf{0.755} & 0.747 & \textbf{0.619} & 0.581 & \textbf{0.619} & 0.607 \\
 SSIM$^t$ & \textbf{0.709} & 0.706 & 0.626 & \textbf{0.645} & \textbf{0.698} & 0.691 & \textbf{0.698} & \textbf{0.698} \\
 PSNR$^h$ & \textbf{17.96} & 17.18 & 22.54 & \textbf{23.70} & \textbf{14.77} & 14.43 & 17.76 & \textbf{17.90} \\
 PSNR$^t$ & \textbf{18.74} & 17.83 & 20.96 & \textbf{21.92} & \textbf{17.56} & 16.83 & \textbf{21.16} & 21.02
\end{tabularx}
\end{table}

\begin{table}[th!]
\centering
\caption{Image quality results for training models by combining $\mathcal{L}_1^{M}$ with $\mathcal{L}_{adv}$, $\mathcal{L}_{SSIM}$, $\mathcal{L}_{VGG}$, or $\mathcal{L}_{Freq}$. Bold text highlights the overall best score for each metric and each image sequence.}\label{tab2}
\begin{tabularx}{0.9\linewidth}{Yc|c|c|c}
\hline
\multicolumn{1}{l}{}                & \multicolumn{4}{c}{T1-synN} \\ \hline
\diagbox{Metric}{Loss func.} & $\mathcal{L}_1^{M} + \mathcal{L}_{adv}$ & $\mathcal{L}_1^{M} + \mathcal{L}_{SSIM}$ & $\mathcal{L}_1^{M} + \mathcal{L}_{VGG}$ & $\mathcal{L}_1^{M} + \mathcal{L}_{Freq}$ \\
\hline\hline
 SSIM$^h$ & 0.701 & 0.729 & 0.736 & \textbf{0.742} \\
 SSIM$^t$ & 0.693 & 0.698 & \textbf{0.712} &  0.710 \\
 PSNR$^h$ & 15.50 & 16.60 & 16.29 &  \textbf{17.72} \\
 PSNR$^t$ & 16.05 & 17.09 & 16.79 &  \textbf{17.98} \\ \hline \hline
\multicolumn{1}{l}{}                & \multicolumn{4}{c}{T1-synC} \\ \hline
 SSIM$^h$ &  0.747 & 0.759 & 0.756 & \textbf{0.768}  \\
 SSIM$^t$ & 0.644 & \textbf{0.649} & 0.643 & 0.648  \\
 PSNR$^h$ & 23.20 & 23.12 & 23.33 & \textbf{23.92}  \\
 PSNR$^t$ & 21.67 & 21.38 & 21.38 & \textbf{22.04 } \\ \hline \hline
 \multicolumn{1}{l}{}                & \multicolumn{4}{c}{T2-synW} \\ \hline
  SSIM$^h$ & 0.591 & 0.607 & 0.616 & \textbf{0.630} \\
 SSIM$^t$ & 0.698 & 0.699 & 0.710 & \textbf{0.717} \\
 PSNR$^h$ & 13.86 & 13.80 & \textbf{14.25} & 14.14 \\
 PSNR$^t$ & 16.29 & 16.11 & \textbf{16.47} & 16.40 \\ \hline \hline
 \multicolumn{1}{l}{}                & \multicolumn{4}{c}{T2-synF} \\ \hline
  SSIM$^h$ & 0.602 & 0.608 & \textbf{0.615} & 0.607 \\
 SSIM$^t$ & \textbf{0.714} & 0.687 & 0.698 & 0.703 \\
 PSNR$^h$ & \textbf{17.51} & 17.18 & 16.93 & 17.37 \\
 PSNR$^t$ & \textbf{21.13} & 20.62 & 20.70 & 21.04
\end{tabularx}
\end{table}

\subsection{Combined loss and submitted solution}
\begin{figure}[ht]
  \centering
  \includegraphics[width=\linewidth]{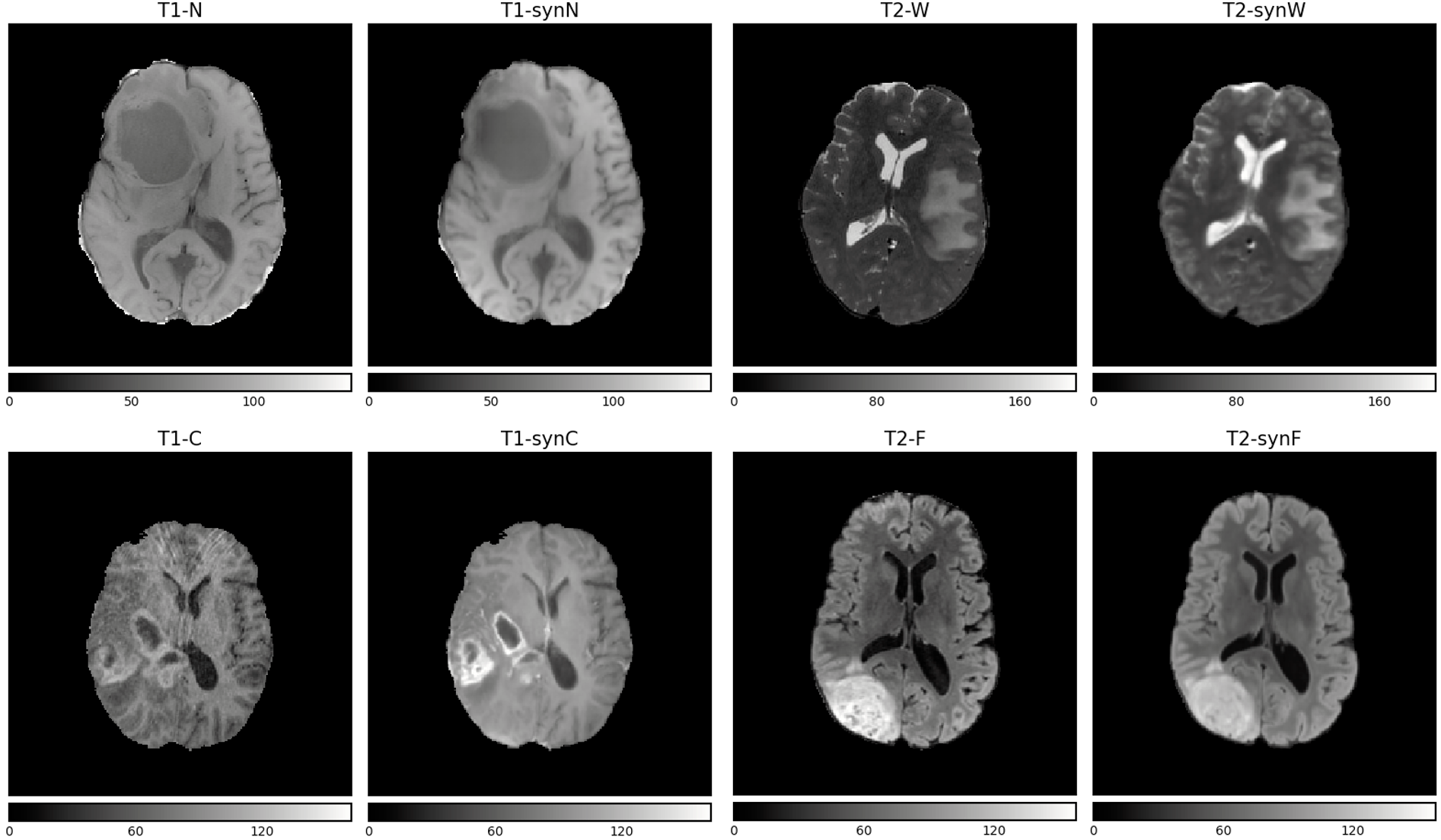}
  \caption{Randomly selected examples from our {\it development} set. Each pair (e.g., T1-N and T1-synN) shows the original image and the corresponding synthetic image. All images are histogram normalized and each pair has the same visualization settings.}\label{fig3}
\end{figure}

We combined all loss functions mentioned in this work and trained four models (i.e., each synthesizing a different sequence) with the results being presented in Table~\ref{tab6}. For our final results, we used tuned weights (based on educated guesses) of each loss term to form the combined loss function. We observe that SSIM$^t$ of T1-synC is generally lower than other sequences, suggesting the difficulty of synthesizing T1-C tumor regions. Visually, this is also confirmed by the randomly selected examples in Figure~\ref{fig3}. T1-synC looks overly bright compared to the original T1-C. However, our model corrected the motion artifacts, which could be advantageous for the downstream segmentation model. For all sequences, we see that $\mathcal{L}_{\text{combined}}$ performed better than the previous results, where we combined two loss functions (cf. Table~\ref{tab2}) or trained with a single loss function (cf. Table~\ref{tab2}).

\begin{table}[th!]
\centering
\caption{Image quality results of synthesis networks for predicting different MRI sequences, i.e., T1-synN, T1-synC, T2-synW, and T2-synF, trained under supervision of a combined loss function.}\label{tab6}
\begin{tabularx}{\linewidth}{Y rr|rr|rr|rr}
\hline
 \multicolumn{1}{l}{} & \multicolumn{2}{c}{T1-synN} & \multicolumn{2}{c}{T1-synC} & \multicolumn{2}{c}{T2-synW} & \multicolumn{2}{c}{T2-synF} \\
\hline
\diagbox{Loss func.}{Metric} & SSIM$^h$ & SSIM$^t$ & SSIM$^h$ & SSIM$^t$ & SSIM$^h$ & SSIM$^t$ & SSIM$^h$ & SSIM$^t$ \\
\hline\hline
$\mathcal{L}_{\text{combined}}$ & 0.816 & 0.771 & 0.761 & 0.656 & 0.754 & 0.790 & 0.694 & 0.733 \\
\end{tabularx}
\end{table}

\subsection{Challenge results}\label{sec:test_results}
\begin{figure}[ht]
  \centering
  \includegraphics[width=0.8\linewidth]{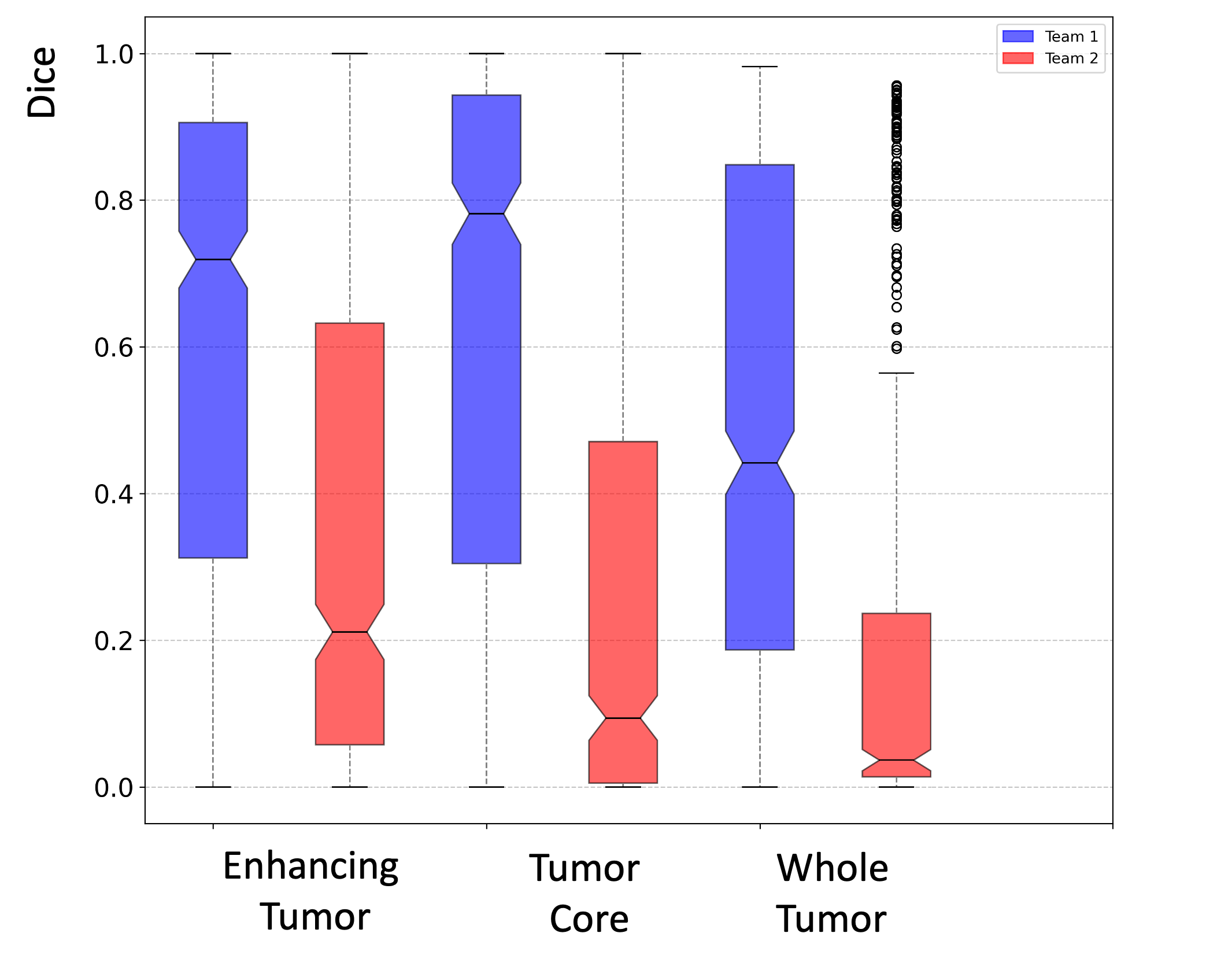}
  \caption{Dice score results for the test set (reported by challenge organizers). Team 1 is our submitted winning solution.}\label{fig2}
\end{figure}

Figure~\ref{fig2} shows the official results of the segmentation method run on the test set with synthetic volumes. The box plot shows the dice score for each class (i.e., Whole Tumor (WT), Tumor Core (TC), and Enhancing Tumor (ET)). Our winning submission is Team 1. We achieved a median dice of 0.72, 0.78, and 0.44 for ET, TC, and WT, respectively. This was significantly better than the other team. Furthermore, our method had an overall mean SSIM of 0.817 for the test set. The challenge organizers did not provide further stratification of the results.

\section{Conclusion}
Most brain MRI segmentation methods are based on the availability of a certain set of MRI sequences and may fail when one of the MRI sequences is absent. The primary goal of the BraSyn 2023 challenge is to encourage the development of MRI sequence synthesis to estimate the missing sequence for downstream tasks of tumor segmentation. In this work, we aimed to investigate the quality of a deep learning-based MRI synthesis model trained under the supervision of different loss functions. Our preliminary results suggest the importance of choosing the right loss function for the synthesis of the MRI sequences. In the future, we plan to provide a comprehensive comparison between individual and combined loss functions on the quality of image synthesis for each MRI sequence. Furthermore, we plan to release an evaluation framework with improved image quality metrics for synthesized medical images to promote good scientific practice.

\bibliographystyle{splncs04}
\bibliography{references}

\end{document}